\documentstyle[11pt,aasms4]{article}
\input{epsf}
\begin{document}

\lefthead{G. Fasano et al.}
\righthead{Early--type Galaxies in the HDF}
 
\title{Early--type Galaxies in the Hubble Deep Field.\\
The $<\mu_e>$--$r_e$ relation and the lack of large galaxies
at high redshift}

\author{Giovanni Fasano}
\affil{Osservatorio Astronomico di Padova, vicolo dell'Osservatorio 5,
35122 Padova, Italy}

\author{Stefano Cristiani}
\affil{Dipartimento di Astronomia, Universit\`a di Padova,
vicolo dell'Osservatorio 5, 35122 Padova, Italy}

\author{Stephane Arnouts}
\affil{Osservatorio Astronomico di Padova, vicolo dell'Osservatorio 5,
35122 Padova, Italy\\
Institut d'Astrophysique de Paris, 98-bis Boulevard Arago, 75014 Paris}

\and
\author{Michele Filippi}
\affil{Dipartimento di Astronomia, Universit\`a di Padova,
vicolo dell'Osservatorio 5, 35122 Padova, Italy}

\begin{abstract}

We present the results of the detailed surface photometry of a sample of
early--type galaxies in the Hubble Deep Field.
Effective radii, surface brightnesses and total $V_{606}$ magnitudes
have been obtained, as well as $U_{300}$, $B_{450}$, $I_{814}$, $J$,
$H$ and $K$ colors, which are compared with the predictions of
chemical-spectrophotometric models of population synthesis.
Spectroscopic redshifts are available for $23$ objects. For other $25$
photometric redshifts are given.  In the $<\mu_e>$--$r_e$
plane the early--type galaxies of the HDF, once the appropriate $K+E$
corrections are applied, turn out to follow the `{\it rest frame}'
Kormendy relation.  This evidence, linked to the dynamical information
gathered by Steidel et al.~(1996), indicates that these
galaxies, even at $z \simeq 2-3$, lie in the Fundamental Plane, in a
virial equilibrium condition.  At the same redshifts a statistically
significant lack of large galaxies (i.e. with $\log r_e^{kpc} > 0.2$)
is observed.

\end{abstract}
 

\section{Introduction}

The formation and evolution of galaxies is a complex process, which
may involve starbursts, infall, interactions (e.g., Dressler et al.\
1994a,b; Lilly et al.\ 1995; Moore et al.\ 1996). The Tully-Fisher
(1977) and the Fundamental Plane (FP) (Dressler et al.\ 1987a;
Djorgovski \& Davis 1987) relations are valuable sources of
information for disk and elliptical galaxies,
respectively. In particular, the FP and its projection on the average
surface~brightness ($<\mu_e>$) effective~radius ($r_e$) plane (Kormendy
relation, Kormendy 1985) have been used as powerful tools for
estimating distances of galaxy clusters (Dressler et al.~1987b,
Lynden-Bell et al.~1988, Faber et al.~1989) and, more recently, for
investigating the evolution of elliptical galaxies (Bender et
al.~1992, Pahre and Djorgovski~1996, Fasano et al.~1997).

If the luminosity profiles and the dynamical structures of the
galaxies are similar, then the very existence of the FP implies a
virial equilibrium condition. Moreover, the slope of the
FP (which slightly differs from that expected in the case of `pure'
virial equilibrium) could imply that the $M/L$ ratio depends on the
mass of the galaxy (Djorgovski and Santiago 1993). Recent advances in
the sensitivity and resolution of the observations both in imaging and
spectroscopy with the {\sl Hubble Space Telescope} (HST) and from the
ground have greatly enlarged the horizon of morphological and
dynamical studies, thus allowing the investigation of the FP in clusters
with redshift up to $z\sim 0.6$ (Kelson et al.~1997). The backward
evolution of the surface brightness derived from the zero point of the
FP in these clusters seems consistent with a passive evolution
of galaxies (Bender et al.~1997). On the other hand, 
it is still matter of debate whether the
slope of the FP (i.e. the function $M/L=f(M)$) varies with
the redshift (van~Dokkum and Franx~1996, Kelson et al.~1997, J\o rgensen
and Hjorth~1997). The difficulty of obtaining reliable velocity
dispersion estimates represents the key obstacle in trying to improve
the results of FP studies, as well as to extend them to higher
redshifts. In spite of its larger scatter, the Kormendy relation 
$<\mu_e>-r_e$ allows to overcome this limitation (Kj\ae rgaard et al.~1993,
Pahre et al.~1996, Fasano et al.~1997). Actually, from deep HST imaging 
of fields around powerful radio-sources and radio-loud quasars, Dickinson 
(1995, 1997a) was able to obtain the Kormendy relation of clusters up
to $z\sim 1.2$. 

The Hubble Deep Field (Williams et al.~1996, see also Ellis 1997; hereafter
HDF) is perhaps the most impressive example of
the recent advances in optical imaging.  The very long
exposure time, the observational procedures and the data handling
techniques of HDF have provided exceptional spatial resolution and
depth.  The unprecedented low surface brightness level reached by the
HDF data turns out to be particularly suited for performing detailed
surface photometry of objects whose brightness slowly decreases
towards the outer regions, as in the case of elliptical galaxies.

We have undertaken a long--term project aimed at producing detailed
surface photometry in the four available bands of a sample of
early--type galaxies in the HDF. The main scientific goal of this
project is to put strong observational constraints to the theories
modeling the evolution of elliptical galaxies (e.g. Tantalo et al.
1996, Kauffmann and Charlot 1997).

The detailed surface photometry of our galaxy sample in the $V_{606}$
band is presented elsewhere (Fasano and Filippi 1998; hereafter FF98). In
the present paper we discuss the improvements that the morphological
information, together with the photometry in different optical and
infrared bands, can produce in understanding the processes of galaxy
formation and evolution.

In Section 2 we report on the database we use to perform the present
analysis, referring to FF98 as far as the original sample selection and
the detailed surface photometry are concerned. In the same section we
describe of the auxiliary data (taken from both network and
recent literature) useful for the present analysis. In Section 3 the
technique we used to extract from the luminosity profiles the relevant
photometric parameters is discussed. In Section 4 we analyze the
relation $<\mu_e>-r_e$ (Kormendy relation; hereafter $KR$) for 
early--type galaxies in the $HDF$. In the same section we discuss the
$K+$evolutionary corrections expected from the current models on the
basis of the available optical and infrared photometry, and derive
photometric redshifts for the galaxies without spectroscopic information. 
Discussion and
conclusions are given in Section 5. Unless otherwise noted we assume 
$H_o=50$ km/s/Mpc and $q_o=0.5$ throughout.

\section{The database and the sample selection}

The present sample of {\it early--type} galaxies has been extracted
from the second release of the $WFPC2-HDF$ frames, in the $V_{606}$
band.  The selection is based on the photometry carried out by Couch
(1996) with the automated SExtractor algorithm (Bertin and Arnouts
1996).

A preliminary list of candidates was defined including all the HDF
objects satisfying the following criteria: {\it a)} Kron {\it AB}
magnitude in the $V_{606}$ band $\le$ 26.3; {\it b)} number of pixels
above the threshold limit of $1.3\sigma$ of the background noise $\ge$
200; {\it c)} star/galaxy classifier $(s/g)\ \le$ 0.6 ($s/g$=1 means
'star'). The combination of the first two limits ensures a sufficient
signal to noise ratio and an adequate number of independent isophotes 
to carry out a reliable morphological analysis. 

Each of the 401 objects matching the above limits was examined with the
{\it IMEXAM--IRAF} tool to produce a first screening against late-type
objects. The preliminary morphological classification was also
compared with the ones by Van den Bergh et al. (1996) and Statler
(1996), finding a general good agreement. Detailed surface photometry
(luminosity and geometrical profiles) was carried out on the resulting
list of 162 {\it early--type} candidates. From the analysis of the 
luminosity and geometrical profiles, 34 objects of the sample have been 
recognized to be '{\it disk--dominated}' objects (likely $Sa$ galaxies), 
whereas 28 more objects showed peculiar or unclassifiable profiles. These 
galaxies were excluded from the final sample of 99 '{\it bona fide}' 
early--type galaxies. 

Further details about the sample selection and surface photometry are
given in FF98.  We recall here that, on the basis of their luminosity
profiles, our $HDF$ {\it early--type} galaxies were divided in three
different classes: (FF98; see also Fasano et al. 1996): {\it 1)} the
'{\it Normal}' class (54 galaxies), for which the de~Vaucouleurs law
is well sampled up to the innermost isophote that is not significantly
affected by the $PSF$; {\it 2)} the '{\it Flat}' class (36 galaxies),
characterized by an inward flattening of the luminosity profiles (with
respect to the de~Vaucouleurs' law) which cannot be ascribed to the
effect of the $PSF$ ; {\it 3)} the '{\it Merger}' class (9 galaxies),
in which isophotal contours show the existence of complex inner
structures (two or more nuclei) embedded inside common envelopes,
roughly obeying the $r^{1/4}$ law.

It is worth to point out that the morphological classification of
galaxies in our sample is necessarily much more a qualitative than a
quantitative one. In most cases the spatial resolution is `critical'
and many faint galaxies appear heavily disturbed by the presence of
close (often multiple) companions or peculiar structures. Sometimes the
only unquestionable characteristic of the object is the presence of a
strong spheroidal component. For this reason, although the main
requirement for final inclusion in our sample is the prominence of the
$r^{1/4}$ component in the luminosity profile, we consider more
appropriate referring to our galaxies as `{\it early-type}'.

An additional problem concerning the morphological classification
comes from the fact that the light distribution of very high redshift
galaxies in the $V_{606}$ band (rest-frame mid and far-IV wavelengths)
may reflect the comparatively chaotic distribution of star-formation
rather than the structural morphology of galaxies. In Section 5 
this point will be addressed in more detail.
  
In the following we will restrict our analysis to the galaxies for
which a spectroscopic redshift is available (23), or a reliable
redshift estimate is possible (25), on the basis of multi--band
(optical {\it and} infrared) photometry (see Section 4.2).

In Table 1 we report the relevant data for this sub--sample: in column
(1) the identification given by Williams et al.~(1996); in columns (2)
and (3) the coordinates; in column (4) the luminosity profile class;
in columns (5) and (6) the effective radius and surface brightness
derived from the $V_{606}$ photometry.

\begin{figure}
\epsffile{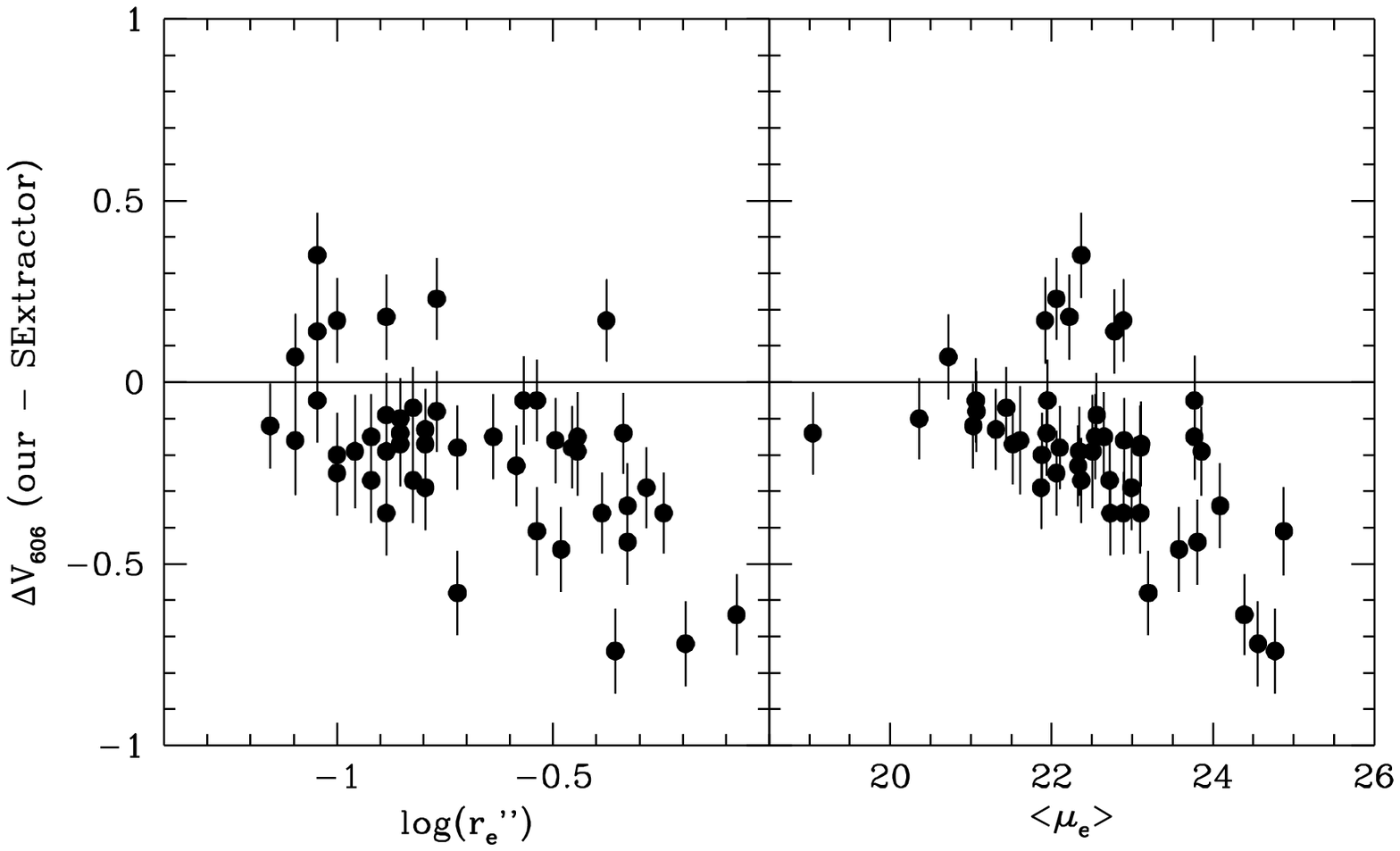}
\vskip -5cm
\caption{
Difference between the magnitudes obtained with 
detailed surface photometry and the SExtractor aperture magnitudes vs. the 
observed average surface brightness (right panel) and the effective radius
(left panel).
\label{fig:aiapsex}
}
\end{figure}

In column (9) the total $V_{606}$ magnitude, estimated with the
detailed surface photometry of FF98, is reported.  In order to obtain
the total magnitudes in the other three WFPC2 bands we have followed
and indirect procedure. We have obtained SExtractor magnitudes in all
four bands using for each galaxy of the sample an optimized aperture
derived from the {$V_{606}$} image.  Then the SExtractor magnitudes
have been scaled by the offset between the total and the SExtractor
$V_{606}$ magnitudes. This procedure has been adopted because for
galaxies with extended profiles like ellipticals the total magnitude
derived from the detailed surface photometry ($V_{606}$ band) is more
reliable than the ones provided by any automated photometry. As an
example, Fig. \ref{fig:aiapsex} shows the biases of the SExtractor
total magnitudes as a function of the galaxy size and surface
brightness. The color photometry scheme we used is a good
approximation if the galaxies do not have significant color
gradients. This is true for '{\it Normal}', nearby ellipticals, where
the color gradients turn out to be usually lower than few tenths of
magnitude (Franx and Illingworth~1990, Peletier et al.~1990).  In
Section 5 we will see that also star--forming galaxies at intermediate
redshift seem not to show significant color gradients. In any case,
the errors expected as due to possible color gradients should be quite
lower than the systematic uncertainties associated with automated
magnitude estimates (see fig.\ref{fig:aiapsex}). The total $U_{300}$,
$B_{450}$ and $I_{814}$ magnitudes are reported in Table 1 (columns 7,
8 and 10, respectively).

Dickinson (1997b) has provided deep, near infrared ($J$,$H$,$K$) images
of the $HDF$ taken with the $KPNO-IRIM$ camera at the Mayall 4--m
telescope (total exposure time of $\sim$45 hours).  The total $J$, $H$
and $K$ magnitudes, again estimated with the SExtractor algorithm, are
reported in columns (11), (12) and (13), respectively.  Column (14) is
a flag defining the reliability of the infrared photometry.

The redshifts of the galaxies are reported in column (15).  The
photometric redshifts are in parentheses, and their uncertainties are
given in column (16).

\section{Deriving $r_e$ and $<\mu_e>$}

Many galaxies in the present sample show a luminosity profile close to
the instrumental PSF. A restoration procedure is needed to extract
useful morphological information.  Fasano et al.(1997) have shown that
the '{\it Multi--Gaussian Expansion}' deconvolution technique ({\it
EMGDEC}, Bendinelli 1991) gives good results in recovering the '{\it
true}' equivalent half-light radius $r_e$ and the corresponding
average surface brightness $<\mu_e>$ from ground--based observations
of elliptical galaxies in distant clusters. In particular, by using
numerical simulations of synthetic elliptical galaxies, Fasano et
al.(1997) concluded that this deconvolution technique is efficient
down to $r_e^{true} \simeq FWHM$.  The input data of the {\it EMGDEC}
algorithm are the '{\it equivalent}' luminosity profiles of the galaxy
and of the $PSF$, both represented in analytical form by means of
suitable series of gaussians.

In the case of the present sample of $HDF$ galaxies, there are
particular problems inherent to the data: {\it (a)} the $PSF$ profile
is known not to be a decreasing monotonic function of radius. This
implies that, in order to give a good representation of the PSF, we
should use a multi--gaussian expansion with free gaussian centers
(i.e. increased number of parameters); {\it (b)} the $PSF$ profile
exhibits quite extended wings which can in principle affect the total
magnitude estimate if some analytic law is assumed to extrapolate the
luminosity profiles; {\it (c)} the presence of strong diffraction
spikes in the $PSF$ can alter the shape of the isophotes, thus
affecting the profiles especially in case of very bright galactic
nuclei; {\it (d)} the shape of the $PSF$ is well known to depend on
the position inside the different $WFPC$ frames.

Concerning the point {\it (a)}, we checked both the '{\it
free-center}' and the '{\it fixed-center}' multi--gaussian
representation of the $PSF$ for a few typical cases. We found
negligible differences in the deconvolved profiles and no differences
in the effective radii between the two procedures. Since the '{\it
free-center}' approach turns out to be much more time consuming with
respect to the other one, we decided to ignore the characteristic
oscillations of the $PSF$ profile in the multi--gaussian
representation of the $PSF$.

The points {\it (c)} and {\it (d)} can be important when dealing with
two--dimensional problems (modeling and removal of point sources,
Lucy deconvolution, etc.). They can be neglected in our case, since
the {\it EMGDEC} deconvolution operates on '{\it equivalent}'
(radially averaged) profiles.

The point {\it (b)} is a delicate one, since it becomes more and more
important as the galaxy size becomes comparable with the $PSF$
size. In this case it is necessary to impose that the multi--gaussian
extrapolation of the outer galaxy profile does not fall below the
$PSF$ profile itself. To this end we forced the extrapolation of the
luminosity profiles of very small galaxies to converge smoothly
towards the $PSF$ profile at large radii.  Luminosity profiles with an
effective radius larger than three times the $FWHM$ were extrapolated
simply with a de~Vaucouleurs' law (see FF98).

Although preserving the total luminosity, the {\it EMGDEC}
deconvolution is not univocal.  Actually, it depends on the
multi--gaussian representation (and extrapolation) of the profiles, as
well as on the so called '{\it regularization}' technique (see
Bendinelli 1991 for details). However, we found that changing these
assumptions within wide ranges produces only marginal changes in the
resulting deconvolved profiles and no significant changes at all in
the corresponding half--light ({\it effective}) radii $r_e$. The
deconvolution turns out to be quite robust as far as the estimation of
the equivalent effective radius $r_e$ is concerned.  It is also worth
stressing that the morphological classification (and then the
inclusion in the sample) is decided before deconvolution. In other
words, the profile shape uncertainties associated with the {\it
EMGDEC} technique do not affect the sample selection.

\begin{figure}
\epsffile{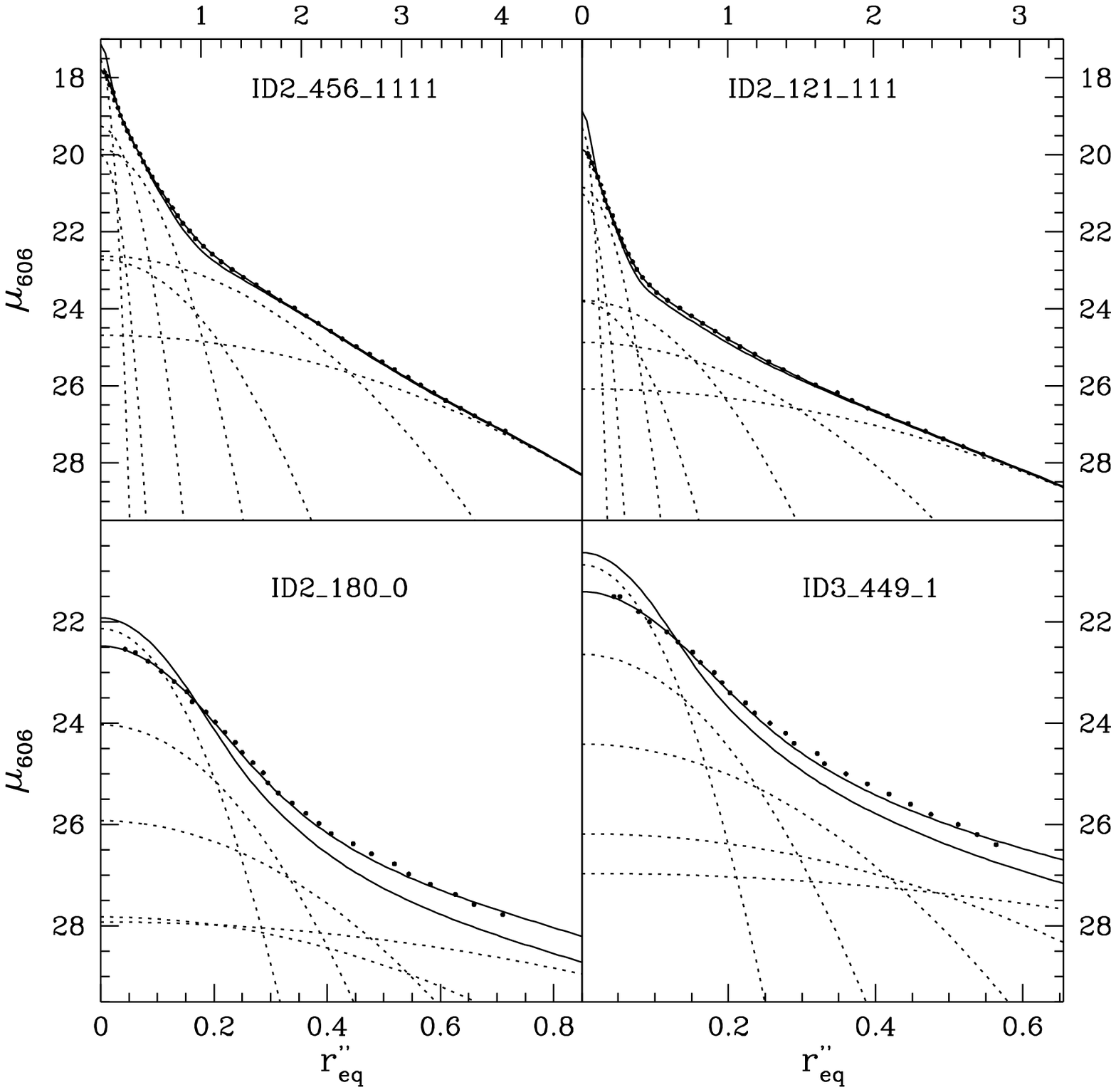}
\vskip -5cm
\caption{
Examples of luminosity profile deconvolution of large (upper panels)
and small (lower panels) HDF ellipticals. The solid lines 
closely following the observed profiles (small dots) represent the
re--convolutions of the deconvolved profiles (solid lines with brighter 
luminosity peaks). The dashed lines in each panel illustrate the 
multigaussian decomposition of the deconvolved profiles.
\label{fig:decex}
}
\end{figure}

\begin{figure}
\epsffile{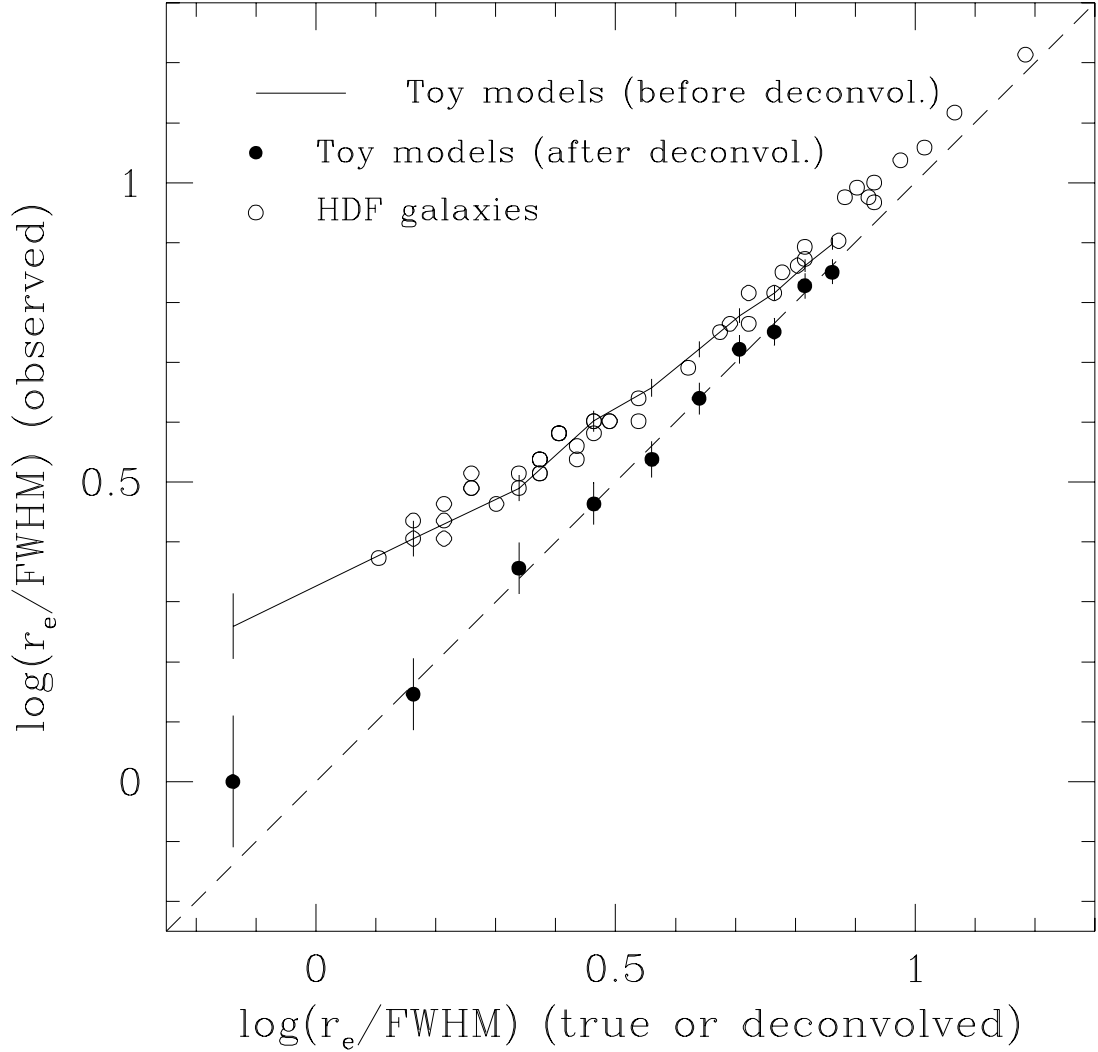}
\vskip -7.5cm\caption
{
Comparison between the effective radii before and after deconvolution
for the HDF ellipticals in our sample (open circles) and for a $10\times 10$
grid of toy models of different sizes and luminosities. Each toy 
model was obtained by adding the proper noise to a synthetic elliptical 
galaxy ($r^{1/4}$ luminosity profile) convolved with the HDF $PSF$. For each
value of the `{\it true}' effective radius in the grid, 10 toy models
with different luminosities were produced and each toy model was processed 
in the same way as the HDF ellipticals (detailed surface photometry and
deconvolution). The solid line connects the average values of $r_e$ 
measured before the deconvolution as a function of the `{\it true}' $r_e$.
The filled circles represent the corresponding average values of $r_e$ 
obtained after the deconvolution. Error bars are reported in both cases.
\label{fig:decef}
}
\end{figure}

Fig. \ref{fig:decex} shows the application of this algorithm in four
extreme cases. The comparison between the effective radii before and
after deconvolution is illustrated in Figure \ref{fig:decef}.

In order to explore how reliable is the {\it EMGDEC} technique in our
particular case, we produced a set of simulations specifically
designed to reproduce the observational conditions of our sample of
$HDF$ ellipticals ($FWHM$/pixel--size ratio, noise, range of
$r_e$/$FWHM$ ratios, range of magnitudes). Several frames of synthetic
elliptical galaxies, with different magnitudes and effective radii,
were first convolved with the theoretical $PSF$ of the $WFPC2$, and
then were processed using our techniques for the extraction and
deconvolution of luminosity profiles.  Figure \ref{fig:decef}
illustrates the results of this analysis. In particular, the bias in
the $r_e$ measured in the simulations (solid line) matches the
distribution of the observed $HDF$ ellipticals (open circles).  Figure
\ref{fig:decef} confirms the previous finding by Fasano et al.(1997)
that the `{\it true}' effective radius can be confidently recovered
down to $r_e/FWHM \simeq 1$ (see filled circles).

In order to obtain reliable error estimates for $r_e$, we have also to
take into account the uncertainties in the total $V_{606}$
magnitudes. Adopting the errors provided in this band by the automated
SExtractor photometry ($\sim 0^m.1$ on average; $\sim 0^m.35$ in the
worst cases), as a conservative estimate for our detailed surface
photometry, we can estimate the log--error in $r_e$ to be $\sim 0.02$
on average ($\sim 0.06$ in the worst cases).  The radius $r_e$ and the
average {\it effective} surface brightness $<\mu_e>$, listed in
columns 5 and 6 of Table 1, are values estimated in the $V_{606}$ band
after deconvolution.

\section{The Kormendy relation of $HDF$ ellipticals}

The relation between the effective radius $r_e$ and the corresponding
average surface brightness $<\mu_e>$ (Kormendy 1985) is just a
projection of the Fundamental Plane of elliptical galaxies (Djorgovski
\& Davis 1987).

The zero point of $<\mu_e>$ in the Kormendy relation (hereafter $KR$)
has been recently used in galaxy clusters at different redshifts, as a
standard candle to perform the classical Tolman test for the expansion
of the Universe and for testing the evolutionary models of 
galaxies (Pahre et al. 1996, Fasano et al. 1997).

In the deep fields (in particular in the $HDF$) we are dealing with a
completely different situation: most of the galaxies are not in
clusters and have very different redshifts. Still, putting the
galaxies in the ($<\mu_e>$--$r_e$) plane is an interesting
exercise, as we will show in the following.

\begin{figure}
\epsffile{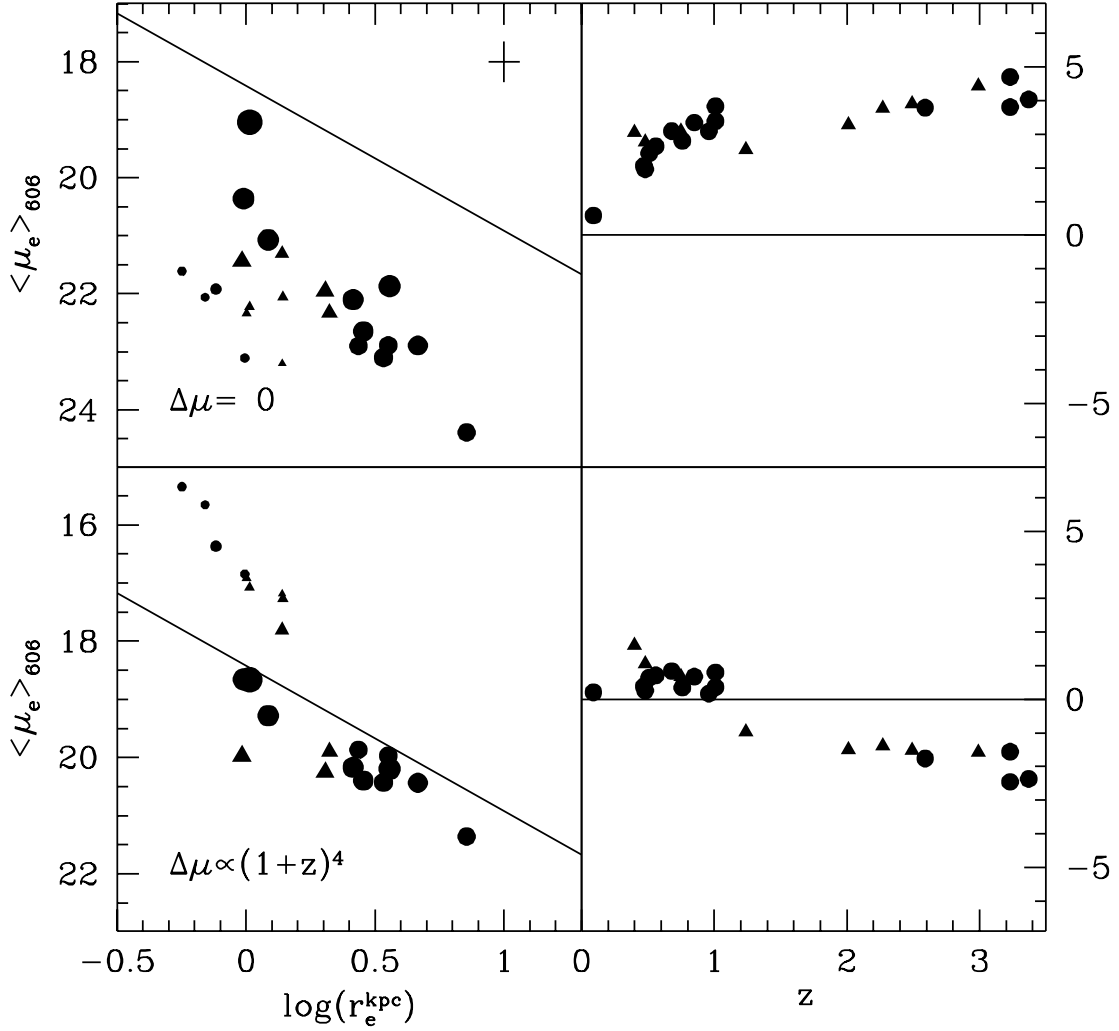}
\vskip -6cm
\caption{
The $\mu_e$--$r_e$ plane for HDF ellipticals with spectroscopic
redshift (left panels). In the upper panel we plot the
$<\mu_e>$ as derived from the deconvolved profiles. The average error
bars in both coordinates are reported in the upper--right part of the
plot. In the lower panel we have applied to $<\mu_e>$ the standard
correction for the cosmological dimming, $(1+z)^{4}$. The solid
line represents the local $KR$ in the $V_{606}$ band. The triangles
represent the galaxies belonging to the classes 2 or 3 of luminosity
profiles.  The size of the symbols decreases with increasing 
redshift.  The panels on the right show the residuals with respect to
the local $KR$ as a function of the redshift.
\label{fig:kor1}
}
\end{figure}

Figure \ref{fig:kor1} shows the $<\mu_e>$--$r_e$ plane for the 23
$HDF$ early--type galaxies in our sample with spectroscopic redshift. The
effective radii (in kpcs) are computed from the deconvolved luminosity
profiles.  The triangles in the plots represent the galaxies belonging
to the above defined `{\it Flat}' and `{\it Merger}' classes of
luminosity profiles (see Section 2). In the left panels the solid
lines represent the `{\it local}' Kormendy relation in the $V_{606}$
band.  It was obtained from the $<\mu_e>$--$r_e$ relation, derived in
the Gunn--$r$ band using the data of J{\o}rgensen et al.(1995), by
shifting the zero point according to the transformations given by
J{\o}rgensen (1994) and Holtzman et al. (1995).  The size of the
symbols in the left panels decreases with increasing redshift.

In the upper panel we plot the $<\mu_e>$ as derived from the
deconvolved profiles. In the lower panel we have applied to $<\mu_e>$
the standard correction due to the cosmological dimming, $(1+z)^{4}$.
The panels on the right show the residuals with respect to the local
$KR$, as a function of the redshift.

A lack of spectroscopic redshifts in the range $1\lesssim z \lesssim
2$ is apparent.  We also note the tendency of high redshift galaxies
($z\gtrsim 2$) to be intrinsically smaller with respect to galaxies
with $z\lesssim 1$.  By applying only the $(1+z)^{4}$ cosmological
dimming correction (lower--left panel in Figure \ref{fig:kor1}) we
obtained two distinct sequences for high redshift and low redshift
galaxies in the $<\mu_e>$ - $r_e$ plane. Both sequences are roughly
parallel to the local $KR$, with a small negative shift in the zero
point at low-redshift and a large positive shift at high redshift.  In
the following section we investigate how such shifts can be explained
in terms of evolutionary effects.

\subsection{The K+E--correction: colors and models}

In order to derive more realistic corrections for the observed surface
brightness of the $HDF$ ellipticals, we need to model their evolution.

We have used the chemical-spectrophotometric models of population
synthesis produced by Bressan et al. (1994, $BCF$) and Tantalo et
al. (1996, $TCBF$) to estimate the K+evolutionary-corrections (K+E)
for elliptical galaxies of different mass, age and evolutionary
history.

The $BCF$ models belong to the `{\it closed-box}' class.  The history
of star formation consists of an initial period of activity followed
by quiescence after the onset of the galactic winds, whose duration
depends on the galactic mass, being longer in the high mass galaxies
and shorter in the low-mass ones.  In the following we will refer to a
{\it passive} model, characterized by an early onset of the galactic
winds, and an {\it active} model, in which star formation continues
until late epochs. They are both described in detail in BCF.

$TCBF$ model the spectrophotometric evolution of elliptical galaxies
as governed by the infall scheme. They simulate the collapse of a
galaxy made of two components, the luminous and dark material.  While
the mass of the dark component is supposed to remain constant with
time, the mass of the luminous component is assumed to grow at a
suitable rate.  The effects of galactic winds powered by supernova
explosions and stellar winds from massive, early-type stars are taken
into account.  We will refer to this scenario as the {\it infall}
models.

\begin{figure}
\epsffile{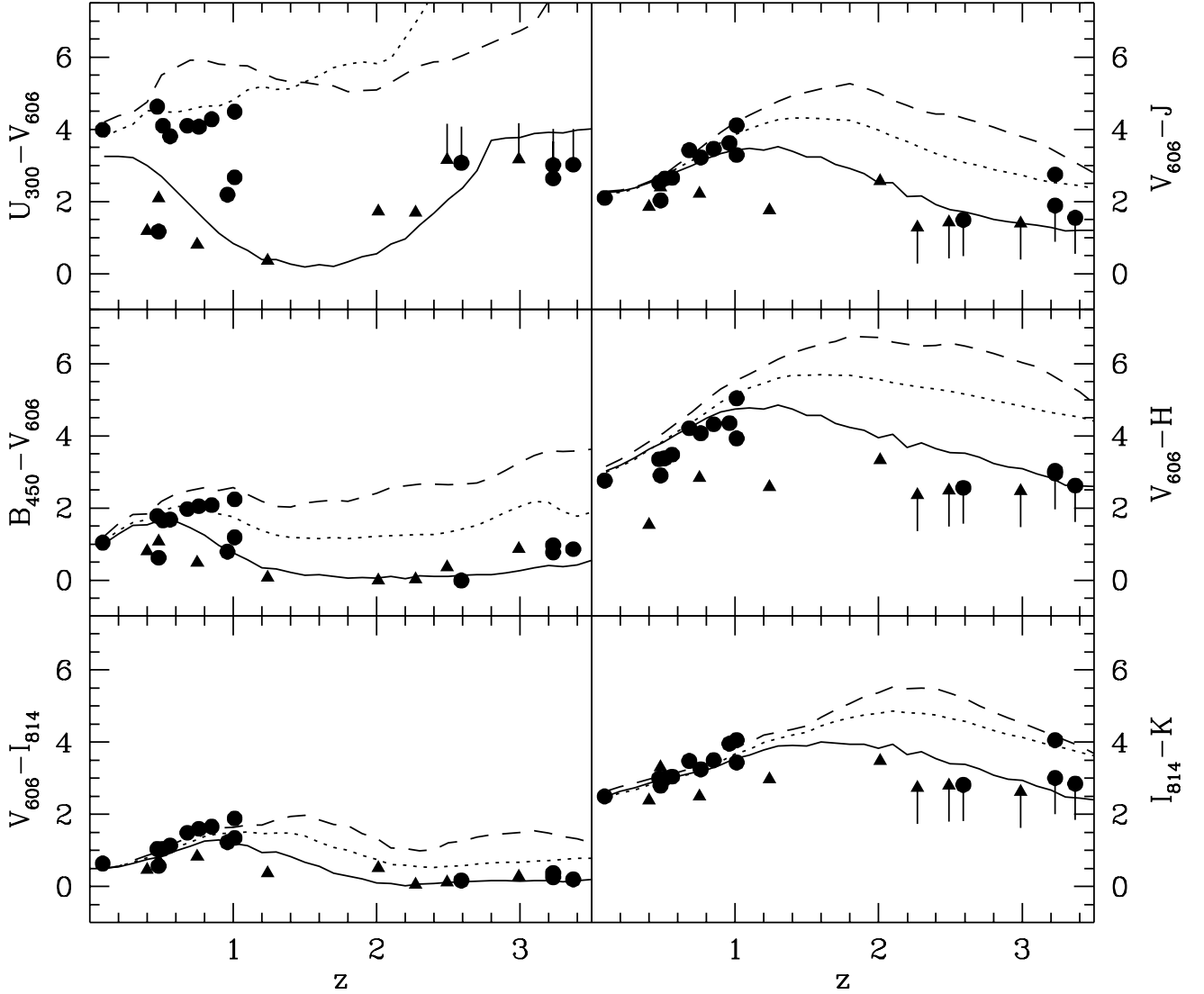}
\vskip -5cm
\caption{
Colors of HDF ellipticals with spectroscopic redshift in our sample as
a function of the redshift. The symbols are the same as in fig. 
\ref{fig:kor1}. The solid, dotted and dashed lines represent the 
BCF--active, BCF--passive and TCBF evolutionary models respectively.
Upper and lower limits are represented by one--sided bars.
\label{fig:kcomp}
}
\end{figure}

In order to verify which subset of the models is able to best describe
the elliptical galaxies observed in $HDF$ we have compared the
observed colors as a function of the (spectroscopic) redshifts with
the model predictions. We have obtained from the models the spectral
energy distributions ($SED$) of a galaxy at a given redshift as a
function of the type of model, galaxy mass, redshift of formation and
cosmological parameter $q_o$. A subset of the results is shown in
Fig.~\ref{fig:kcomp}, where the best--fitting models for each one of
the three evolutionary schemes are compared with the observations. It
can be seen that a model of a $3 \cdot 10^{12} ~ M_\odot$ galaxy with
an `{\it active}' star formation until the age of $11$ Gyr with a
redshift of formation $z_{form} = 5$ in a universe with $q_o = 0.5$
provides a reasonable `global' description of the observed colors.  At
low redshift, two different sequences of galaxy colors are apparent in
the $U-V$ (and in minor degree in the $B-V$).  To fully reproduce
them, different models or at least different choices of the parameters
(e.g. $z_{form}$) characterizing a given model would be required.  We
have not attempted to do it in this simplified approach, since we were
interested only in the K+E corrections to be applied to the $V_{606}$
band.  It is remarkable that all the galaxies belonging to the bluer
population show `{\it abnormal}' luminosity profiles (types 2 and 3 in
Tab. 1, squares in Fig.~\ref{fig:kcomp}).

In general, smaller values of $q_o$ can be compensated (although not
exactly) by a corresponding variation of the redshift of formation, in
the sense that for lower values of $q_o$ the galaxy colors at a given
redshift tend to be reproduced by models with smaller $z_{form}$.

The {\it passive} closed--box model and the models governed by the
infall are clearly not able to reproduce the color--$z$ diagrams with
a unique redshift of formation.

In the present analysis we have not included the effects of dust
which, due to the shape of the extinction law, are potentially more
and more important at increasing redshifts.  In general to reproduce
the observed colors in presence of dust a smaller redshift of
formation is required.  On the other hand, a larger unabsorbed
luminosity has to be assumed and the two effects tend, at least
qualitatively, to compensate in the estimation of the $K+E$
correction.

\begin{figure}
\epsffile{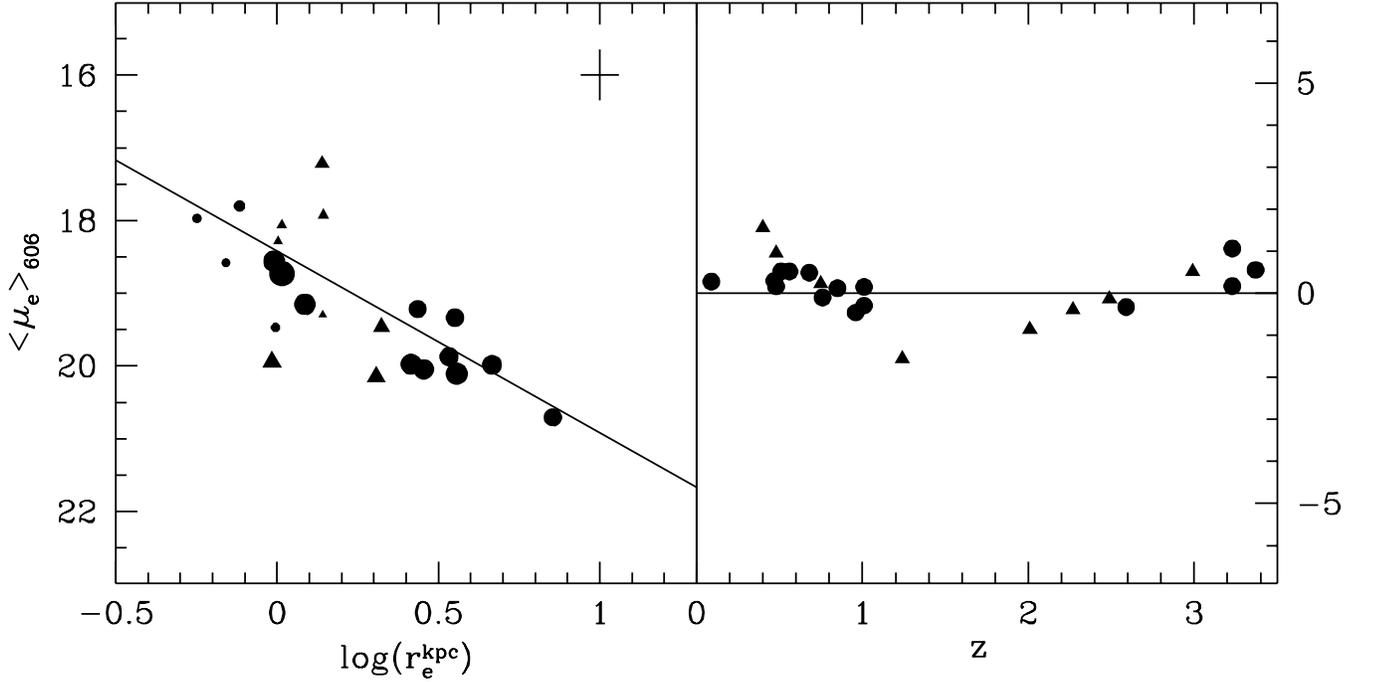}
\vskip -5cm
\caption{
The $\mu_e$--$r_e$ relation for HDF ellipticals with spectroscopic
redshift in our sample (left panel). Symbols are the same as in fig.
\ref{fig:kor1}. In this case we have applied to the observed $<\mu_e>$ 
the K+E-corrections predicted by the $BCF$ {\it active} model with
$q_o=0.5$ and $z_{form}=5$.  The panel on the right shows the
residuals with respect to the local $KR$, as a function of the
redshift.
\label{fig:kor2}
}
\end{figure}

If we apply to the observed $<\mu_e>$ the K+E-corrections predicted by
the $BCF$ {\it active} model with $q_o=0.5$ and $z_{form}=5$, we find
the K--relation shown in Figure \ref{fig:kor2}, which for all
redshifts is in fairly good agreement with the local relation.  Again,
we note that all galaxies with very high redshift (small symbols) are
confined in the high surface brightness, small size region of the
plane.

\subsection{Extending the sample: photometric redshifts}

In order to improve the statistics, we have estimated the photometric
redshifts of the 25 galaxies in the present sample having sufficient
and reliable infrared information (i.e. detection in $J$, $H$ and $K$,
no deblending problems), besides the optical magnitudes.

The photometric redshift technique has been successfully used by
various authors in the $HDF$ (e.g. Lanzetta et al. 1996) and in other
fields (e.g. Connolly et al. 1995, Giallongo et al. 1997).

We have adopted an approach of the type described by Giallongo et
al.(1997), computing a large number of galaxy SEDs as a function of
the formation redshift, redshift of the observation, internal
extinction and evolutionary scenario and comparing them with the
observed colors.  In the present case only population synthesis models
of elliptical galaxies (those provided by $BCF$ an $TCBF$) were used.
In this way, the morphological information allowed us to restrict the
range of the practicable models, avoiding possible degeneracies.  In
practice, the technique exploits at low redshift the position of the
400 nm break and at high-redshift the drop in the galaxy flux due to
the combination of the IGM plus intrinsic Lyman absorption.

\begin{figure}
\epsffile{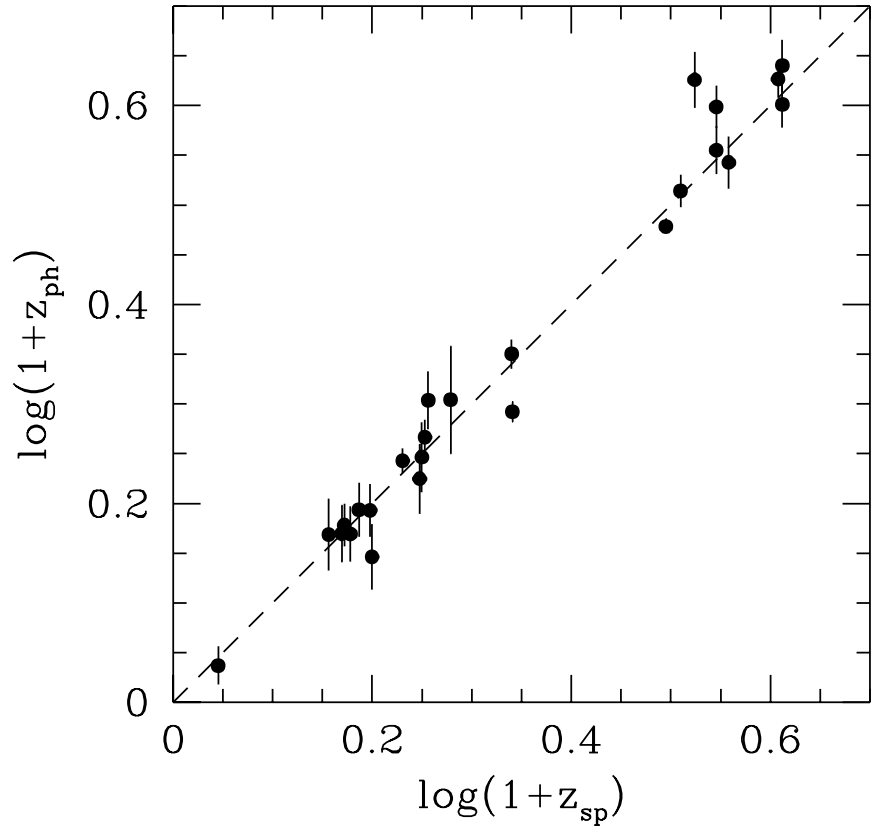}
\vskip -5cm
\caption{
Comparison between the spectroscopic and photometric redshifts.
\label{fig:dz}
}
\end{figure}

A complete discussion of the `{\it photometric redshift}' procedure is
beyond the scope of this paper and will be given elsewhere (Arnouts et
al. in preparation).  In Fig. \ref{fig:dz} we show the comparison
between the spectroscopic and photometric redshifts for the 23
galaxies in our sample having both redshift determinations.

Fig.~\ref{fig:kor3} shows the $KR$ we obtain for the whole sample,
including the galaxies with photometric redshifts (open symbols with
error bars) and those with spectroscopic redshifts (same symbols as in
Fig.~\ref{fig:kor2}).  Again the triangles represent the galaxies with
`{\it abnormal}' luminosity profiles and the size of the symbols is
decreasing with increasing redshift.

The $KR$ for galaxies with photometric redshift estimates shows a
larger dispersion with respect to the spectroscopic sample. However,
the general trend is preserved and the 2--D Kolmogorov--Smirnov test
(Fasano and Franceschini 1987) cannot reject the null hypothesis that
the two distributions are drawn from the same parent population with
at more than $82\%$ significance level. In any case different kinds of
biases are expected to affect the two samples (e.g.  objects with
spectroscopic identifications tend to have on average brighter
magnitudes and higher central surface brightnesses).

\section{Discussion and Conclusions}

\begin{figure}
\epsffile{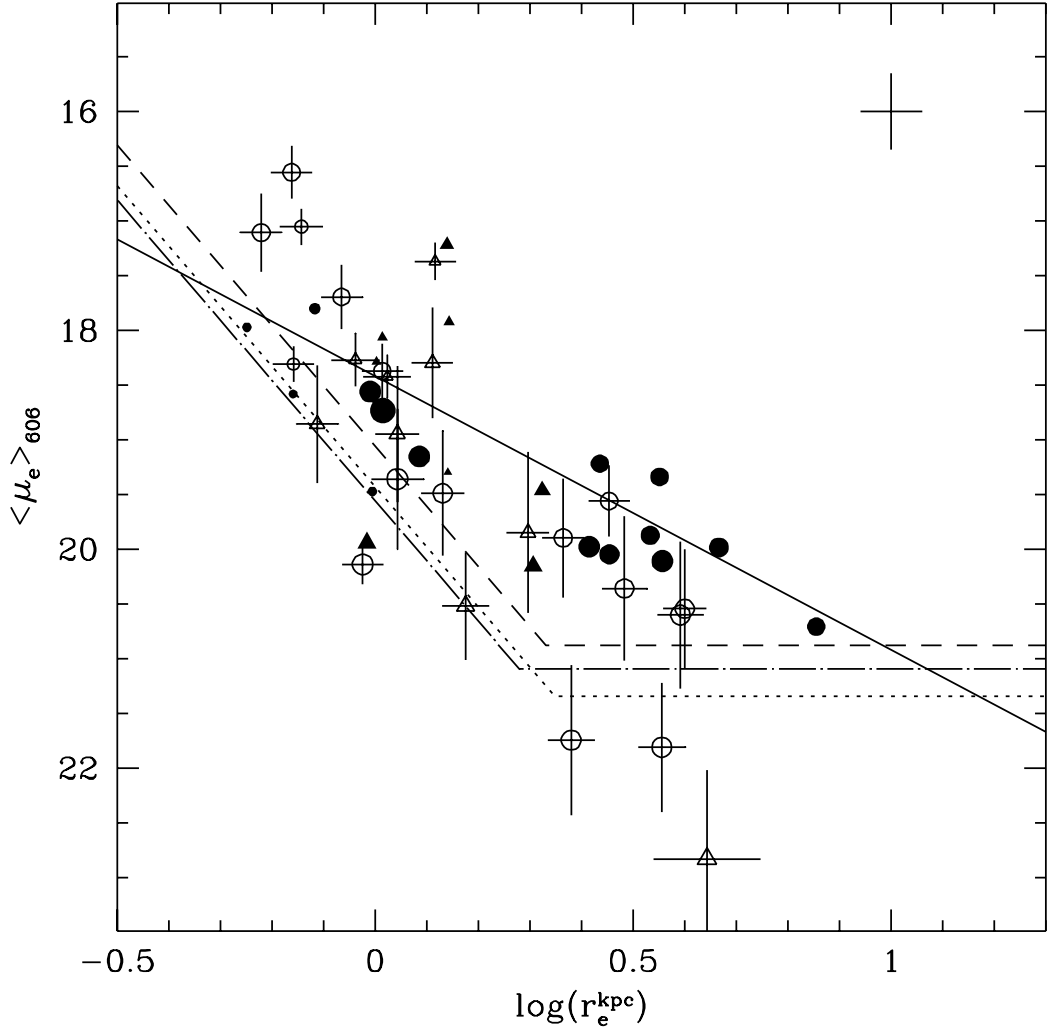}
\vskip -6cm
\caption{
The Kormendy relation for HDF ellipticals in our sample. The filled and
empty symbols refer to galaxies with spectroscopic and photometric
redshift, respectively. The symbols and the symbol sizes are the same
as in fig. \ref{fig:kor1}. The error bars of galaxies with photometric 
redshift are due to the uncertainty in the redshift estimate, as well as 
to the average errors of $r_e$ and $<\mu_e>$. In the framework of the $BCF$
active models, the regions below the dotted, dashed and dot--dashed lines 
represent the regions of the Kormendy plane which are '{\it forbidden} because 
of the adopted selection criteria, for z = 1, 2 and 3, respectively.
\label{fig:kor3}
}
\end{figure}

Figure \ref{fig:kor3} indicates that, if we apply to the observed
$<\mu_e>$ the K+E-correction deduced from the galaxy models that
properly reproduce the observed colors, the early--type galaxies in the
$HDF$ follow the local Kormendy relation at any redshift.

The Kormendy relation is well known to be a projection of the
Fundamental Plane (FP) which roughly represents the virial equilibrium
condition of galaxies, except for a small tilt interpreted in terms of
an $M/L \propto M^{\alpha}$ relation (Faber et al. 1987).

Steidel et al. (1996) measured the line widths of compact spheroidal
galaxies at $z > 3$ which, if interpreted as due to gravitational
motions, imply velocity dispersions of the order $180 \le \sigma \le
320$ km s$^{-1}$.  Such values would place the high-z galaxies of the
present sample in the typical region of the Fundamental Plane
corresponding to early--type galaxies, showing that, even at $z>2$, 
these galaxies obey an average, rest--frame, virial equilibrium condition.
It is worth noticing that, even assuming a significant non-virial contribution 
(e.g. stellar winds, SN ejecta) to the observed line widths,
the high-z galaxies would still be in the FP region occupied by
intermediate-luminosity elliptical galaxies.

Fig.~\ref{fig:kor3} also shows that, high redshift galaxies are
confined in the high surface brightness, small effective radius region
of the $KR$ plane.  Is this tendency due to the fact that large
ellipticals have a faint surface brightness (according to the Kormendy
relation itself) ?  Large galaxies at high redshifts could lie below
the detection limit.  We have checked at various redshifts what
regions of the Kormendy plane result `{\it forbidden}' because of the
adopted selection criteria.  The regions of Fig.~\ref{fig:kor3} below
the dotted, dashed and dot--dashed lines represent, in the framework
of the $BCF$ active models, the forbidden regions of the Kormendy
plane for z = 1, 2 and 3, respectively.  We conclude that, had there
been large galaxies at high redshifts in the $HDF$, they would have
been included in our sample.

However, it remains to be demonstrated that the comoving volume
density of large high redshift galaxies is high enough to ensure the
presence of a significant number of them in our sample.  We have
estimated in two different ways the expected fraction, $F_L$, of
$z>1.5$ galaxies with $\log(r_e^{kpc}) > 0.2$.

\begin{enumerate}

\item 
Assuming that the $r_e$ distribution at $z>1.5$ is the same observed
in the Coma cluster (Lobo et al. 1997) for galaxies in the same
absolute magnitude range, we found $F_L = (166/299) \simeq 1/2$.  To
carry out this computation, we applied the $K+E$ corrections of the
BCF active models and transformed the isophotal radii (at
$\mu_V=26.5$) measured by Lobo et al. (1997) into effective radii by
using the average growth curve of the elliptical galaxies given by
de~Vaucouleurs et al. (1976).  It should be noted that the sample by
Lobo et al. (1997) contains all morphological types and the
distribution of $r_e$, although ellipticals are preponderant in Coma,
may be affected by the presence of spirals and irregulars.

\item 
In an alternative approach we assumed that the galaxies at $z>1.5$ -
again with the $K+E$ correction derived by the BCF active models -
follow a Schecter-type luminosity function with an $M_B^*= -21.7$
(Ferguson and Sandage 1991).  Assuming that the luminosity profiles
follow the de~Vaucouleurs law, we can use the local Kormendy relation
to transform the distribution of absolute luminosities in a
distribution of effective radii, according to the formula: $
-2.5~\log{r_e}= (M_{606} + 20.15) $. In this way we found again $F_L
\simeq 1/2$.

\end{enumerate}

Although both estimates are based on rough approximations, they turn
out to be robust with respect to changes of the models adopted for the
$K+E$ corrections, the luminosity function parameterization and
possible contaminations of the diameter and/or luminosity distribution
functions.  Therefore we conclude that at $z>1.5$ a scarcity of large
early-type galaxies is observed with respect to the local
universe.

What type of selection effects may affect the present sample ?

The choice of the $V_{606}$ as the selection band implies a bias at
redshifts $z>1$ in favor of star-forming galaxies.  So it is not
surprising that at $z>2$ the observed colors of the galaxies favor the
`active' models. Galaxies following a `passive' or `infall' evolution,
if formed at sufficiently large redshifts, quickly become too faint to
be observable. So, a population of redder galaxies which had their
major burst of star formation few e-folding times before the epoch of
the observations would be undetected in the HDF.

Is also our perspective of the morphology of galaxies
possibly biased by the fact that at $z>1$ the $V_{606}$ band is
sampling the ultraviolet part of the SED ?  The UV morphological
properties of nearby elliptical galaxies are known to match those in
the optical (see O'Connell 1997 and references therein), but for these
systems the UV light is coming from evolved stars (the UV upturn).  On
the contrary, in interpreting the morphology of star-forming
galaxies at substantial redshifts, one must bear in mind that moving to
UV wavelengths the light becomes proportional to the star-formation
rates and not to the global stellar content of a galaxy.  In principle
the UV profiles could depart from an $r^{1/4}$ law, reflecting the
comparatively more chaotic morphology of the star formation rather
than the structural morphology of a galaxy.

\begin{figure}
\epsffile{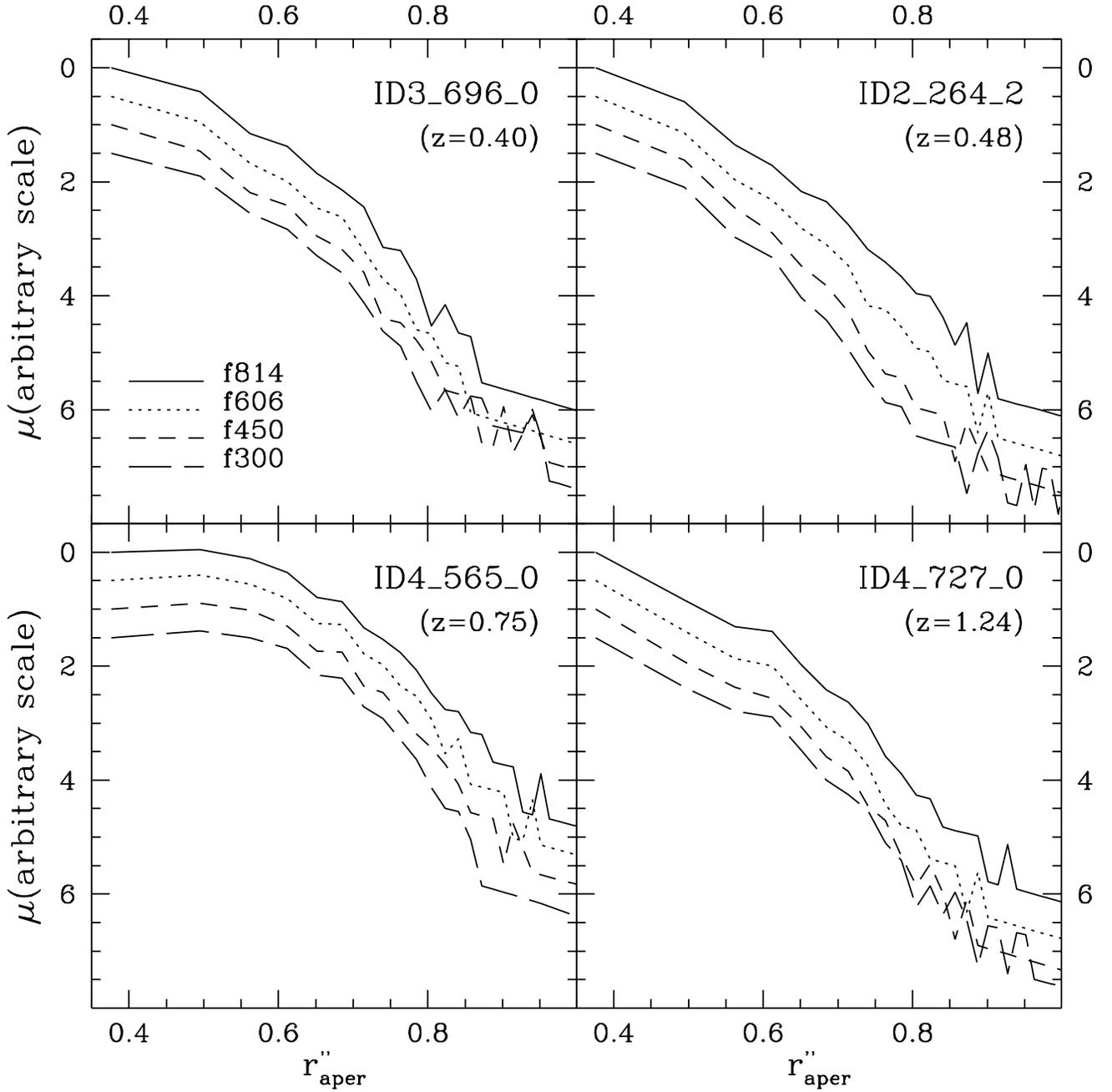}
\vskip -5cm
\caption{
Comparison among the aperture luminosity profiles in the four WFPC bands
for the four star-forming galaxies in our sample with spectroscopic redshift 
$0.4 < z < 1.25$. The surface brightness is expressed in arbitrary units 
and the central values of $\mu$ in the four bands are shifted each other
by $0^m.5$. 
\label{fig:profil}
}
\end{figure}

On the other hand, the SED of unreddened star-forming galaxies at
high-z is essentially flat from $\sim 1200\AA$ to $\sim 4000\AA$ rest
frame, and to even longer wavelengths if the galaxy is young. At the
wavelength of the $V_{606}$ WFPC2 filter, this corresponds to
observing galaxies in the redshift interval $0.5 < z < 4$, with
minimal $K$--corrections (see Giavalisco et al. 1996). This means that
it is not unfair to compare the UV morphologies of the $z > 2$ galaxies
with those of the $z\sim 1$ systems.  In order to check quantitatively
this conclusion we produced, in the four WFPC2 available bands, the
`{\it aperture}' luminosity profiles of the galaxies with
spectroscopic redshifts $z < 1.5$ and $U_{300}-V_{606} < 2$. The
SED of these galaxies ($3-696-0$, $2-264-2$, $4-565-0$ and $4-727-0$),
obtained from the optical and infrared magnitudes in Table 1, clearly
indicates star--forming activity. Nevertheless, the
Fig.~\ref{fig:profil} shows that the light distribution in these
objects is ostensibly the same in a wide range of rest--frame
wavelengths (from the $V$ band to the far-$UV$).

\section{acknowledgments}

We thank A. Bressan, S. Charlot, C. Chiosi and E.V. Held for
enlightening discussions and an anonymous referee for stimulating
comments.  This work was partially supported by the ASI contracts
94-RS-107 and 95-RS-38, by the TMR program of the European Community
and by the Formation and Evolution of Galaxies network set up by the
European Commission under contract ERB FMRX-CT96-086 of its TMR
program.

\vfill\newpage

\end {document}